\documentclass[10pt,letterpaper]{article}

\usepackage{cogsci}
\usepackage{pslatex}
\usepackage{times}
\usepackage{helvet}
\usepackage{courier}
\usepackage{pgfplots}
\usepackage{pgfplotstable}
\usepackage{siunitx}
\usepackage{bm}  
\usepackage{mathrsfs}
\usepackage{color, colortbl}
\usepackage{dcolumn}
\usepackage{multicol}
\usepackage{amsmath}

\definecolor{LightCyan}{rgb}{0.88,1,1}
\definecolor{Green}{rgb}{0,0.5,0.25}
\definecolor{Red}{rgb}{0.5,0,0}

\newcommand{\trb}{\bm{\tau}}
\newcommand{\fub}{\mathrm{\mathbf{f}}}
\newcommand{\ACTM}{\bm{M}}
\newcommand{\Gop}{\mathscr{G}}
\newcommand{\elab}[1]{{\em{#1}}}
\newcommand{\epa}[3]{(EPA:$\{#1,#2,#3\}$)}
\newcommand\minus{%
    \setbox0=\hbox{-}%
    \vcenter{%
        \hrule width\wd0 height \the\fontdimen8\textfont3%
    }%
}

\newcommand\mcb[1]{\multicolumn{1}{c|}{#1}}
\newcommand\mcnb[1]{\multicolumn{1}{c}{#1}}
\newcolumntype{N}{D{.}{.}{-1}}
\newcolumntype{O}{D{,}{,}{2.2}}
\newcolumntype{M}{D{.}{.}{2.1}}
\newcolumntype{L}[1]{>{\raggedright\let\newline\\\arraybackslash\hspace{0pt}}m{#1}}

\newcommand{\mi}{\minus}
\newcommand{\commentout}[1]{}

\newenvironment{para_packed_enum}{
    \begin{changemargin}{0em}{0em}  
        \begin{enumerate}
            \setlength{\itemsep}{1pt}
            \setlength{\parskip}{0pt}
            \setlength{\parsep}{0pt}
            \setlength{\itemindent}{0.9em}
        }{\end{enumerate}
    \end{changemargin}
}

\newenvironment{para_packed_item}{
    \begin{changemargin}{0em}{0em} 
        \begin{itemize}
            \setlength{\itemsep}{1pt}
            \setlength{\parskip}{0pt}
            \setlength{\parsep}{0pt}
            \setlength{\itemindent}{0.9em}
        }{\end{itemize}
    \end{changemargin}
}

\title{Socio-Affective Agents as Models of Human Behaviour in the Networked Prisoner's Dilemma}
 
\author{{\large \bf Joshua D. A. Jung and Jesse Hoey} \\
  \{j35jung, jhoey\}@cs.uwaterloo.ca \\
  David R. Cheriton School of Computer Science, University of Waterloo \\
  Waterloo, Ontario, Canada }

\begin{document}

\maketitle

\begin{abstract}
Affect Control Theory (ACT) is a powerful and general sociological model of human affective interaction. ACT provides an empirically derived mathematical model of culturally shared sentiments as heuristic guides for human decision making. BayesACT, a variant on classical ACT, combines affective reasoning with cognitive (denotative or logical) reasoning as is traditionally found in AI. Bayes\-ACT allows for the creation of agents that are both emotionally guided and goal-directed. In this work, we simulate BayesACT agents in the Iterated Networked Prisoner's Dilemma (INPD), and we show four out of five known properties of human play in INPD are replicated by these socio-affective agents. In particular, we show how the observed human behaviours of network structure invariance, anti-correlation of cooperation and reward, and player type stratification are all clearly emergent properties of the networked BayesACT agents. We further show that decision hyteresis (Moody Conditional Cooperation) is replicated by BayesACT agents in over $2/3$ of the cases we have considered. In contrast, previously used imitation-based agents are only able to replicate one of the five properties. We discuss the implications of these findings in the development of human-agent societies. 

\end{abstract}

\section{Introduction}
A social dilemma may be defined as a ``situation in which individual rationality leads to collective irrationality''~\cite{kollock1998social}, which is to say that agents acting out of purely rational self-interest will decrease the total reward received by all agents. Such situations are ubiquitous in modern society: cutting off a fellow driver, jumping a queue, and accepting under-the-table payment are minor, if not quite innocuous, examples. Of course, not all such conflicts need be small. For instance, we might view carbon emissions as a large scale Tragedy of the Commons dilemma with potentially disastrous consequences. Solutions to social dilemmas can therefore have broad application, and have seen much study by psychologists, sociologists, and more recently, computer scientists.

As a means of abstraction, social dilemmas are frequently cast as normal-form games. In the simplest case, these games involve two players who must simultaneously choose an action. Their choices are then used to assign each a payoff from a pre-defined reward matrix. In general, the players' matrices need not be identical (i.e. they may be rewarded under different circumstances), but must be known by both parties before actions are chosen.

Over decades of research, several canonical dilemmas/games ha\-ve emerged, each framing an otherwise sterile reward matrix within a relatable story. Examples include Assurance, Chicken, Public Goods, and the Tragedy of the Commons. None, however, is better studied than the Prisoner's Dilemma, a game which has produced thousands of studies~\cite{kollock1998social}.
\commentout{
    In its case, one version of the canonical story may be told as follows. The police are holding two partners-in-crime in separate rooms with no way for them to communicate with each other. During their interrogations, each has the option to either defect and snitch on the other, or cooperate with their partner and remain silent. Their choices yield one of the following results.
    \begin{para_packed_item}
        \item If both partners cooperate, the prosecution fails to make a strong case, and each receives a 2 year sentence reduction (the ``cooperation reward'').
        \item If only one partner defects, he receives a 3 year sentence reduction (the ``temptation''), and the other gets no reduction (the ``sucker's payoff'').
        \item If both partners defect, each receives only a 1 year sentence reduction (the ``punishment'').
    \end{para_packed_item}
}
\begin{table}[tb]
    \centering{
        \begin{tabular}{cc}
            \begin{tabular}{c|c|c}
                & C   & D   \\
                \hline
                C & R,R & S,T \\
                \hline
                D & T,S & P,P \\
            \end{tabular}
            &
            \begin{tabular}{c|c|c}
                & C   & D   \\
                \hline
                C & 2,2 & 0,3 \\
                \hline
                D & 3,0 & 1,1 \\
            \end{tabular}
            \\
        \end{tabular}
    }
    \caption{\label{tab:PDGame} The general reward matrix and an example reward matrix for the Prisoner's Dilemma. For all pairs, the row player's reward is shown first, and the column player's second.}
\end{table}

The reward structure of this game is summarized in Table \ref{tab:PDGame}, where cooperation and defection are represented by $C$ and $D$ respectively, $R$ is the reward for cooperation, $S$ is the sucker's payoff for being duped, $T$ is the temptation payoff, and $P$ is the punishment payoff. For a game to be considered a true Prisoner's Dilemma, the following inequality must be obeyed~\cite{kollock1998social}.\footnote{Note that other permutations of this inequality may produce other social dilemmas (eg. the Assurance Game is characterized by $R>T>P>S$) or no dilemma at all (eg. $R>S>T>P$, where cooperation dominates).}

\begin{equation} \label{eq:PDInequality}
T > R > P > S
\end{equation}

As a direct result, for a single play of this game (or any finite number of repeated games between two players) defection is a dominant strategy and there is a Nash equilibrium for pure defection~\cite{kreps1982rational}.  In other words, regardless of the choice the other makes, a player is rewarded more for defection than cooperation. However, it has repeatedly been shown that, in spite of this, human players do not adhere to an all-defect strategy~\cite{flood1958some, deutsch1958trust, grujic2014comparative}.

Humans are not strictly rational creatures. This allows us to pursue a goal that could be viewed as non-traditional within the scope of Artificial Intelligence: rather than attempting to find the "best" strategy that collects the highest reward, we look for one that appears to best model the play of actual people. Previous work in the same vein has analyzed human play~\cite{grujic2014comparative} by comparing human INPD experiments, and found that humans exhibit five major properties.  First, human play is invariant to network structure. Second, global cooperation rates decline over time, but remain non-zero. Third, cooperation is anti-correlated with reward. Fourth, most humans exhibit ``moody conditional cooperative'' behaviour, and fifth, human play is stratified into four major groups.

Our long term goal is to build automated agents that could become authentic and accepted members of a mixed society of biological and technological agents. In order to achieve this goal, we must build agents that understand and respect the emotional or affective reasoning that is a hallmark of human intelligence~\cite{Kahneman11,Damasio1994}. In this paper, we follow recent work in sociological artificial intelligence~\cite{HoeyBACT15,SchroederHoeyRogers2016,asghar2015intelligent} in which agents are built upon the principles of affect control theory (ACT). ACT arises from work on the sociology of human interaction~\cite{Heise2007}, and proposes that social perceptions, behaviours, and emotions are guided by a psychological need to maintain consistency in culturally shared affective sentiments about social situations. This ``affect control principle'' has been shown to be a powerful predictor of human behaviour across a wide range of domains~\cite{MacKinnon2014}. 

In this paper, we investigate whether automated agents built on the principles of ACT can replicate the five properties of human play in the INPD foud by Gruji\'c et al.~\cite{grujic2014comparative}.  We compare the ACT-based agents to agents playing standard imitative strategies~\cite{vilone2014social} across a range of different network structures and payoff matrices. We find that while the imitator agents only replicate one of the human properties, the ACT-based agents clearly replicate four out of five, and somewhat replicate a fourth. The only property that the ACT-based agents did not reliably replicate was the decline of cooperation rates over time. ACT-based agents remained much more cooperative in the long-term than humans, a property that was also observed in the human studies reported in~\cite{jung2016grounding}.



To achieve our long-term goal of acceptable technological agents in human society, we wish to produce an agent that can act as a human analogue in the context of games or game-like situations, and we would like it to determine its actions in a way that is as consistent with human cognition as is possible. Meeting this condition grants predictive power, as many important activities may be cast as games, (e.g. auctions and elections). Additionally, it has been shown (e.g. by Kiesler et al.~\cite{kiesler1996prisoner}) that humans behave differently when they know that they are playing against a machine. In many applications, this is undesirable. An agent that is better at playing like humans is therefore better at playing with humans.

In this paper, we study one particular class of social dilemmas that are known to have many analogies in human society.  Although the game-like structures we are studying are limiting in that they do not directly involve humans in real-life situations,
the basis of our work in well-established, empirically derived sociological models of human behaviour helps substantiate the generalisability of our results.

\section{Related Work}

\subsection{Affect Control Theory and BayesACT}
ACT, as first formalized by David Heise \cite{heise1977social}, begins from the well-established fact that every concept carries with it a culturally shared emotional sentiment \cite{osgood1952nature}. That is to say, two individuals within a close social group will tend to make similar emotional evaluations of a given thing (car, dog, hospital, etc.), action (hug, kick, kill, etc.), or person (banker, child, senior, etc.).\footnote{Note that there is no hard cut-off for what constitutes a ``close social group''. In practice, a country may constitute such a group, as has been the case for the USA~\cite{francis2006mean}.} Further, ACT posits that this identity can be measured with three numerical values $\in [-4.3, 4.3]$\cite{heise1977social}:~\footnote{The scale of $4.3$ is a historical convention.}
\begin{para_packed_item}
    \item {\bf E}valuation: good/nice  vs. bad/awful
    \item {\bf P}otency/Power: strong/big/powerful vs. weak/little/po\-werless
    \item {\bf A}ctivity: exciting/active/fast/loud vs. calm/passive/sl\-ow/quiet
\end{para_packed_item}
The EPA space is hypothesized to be a universal organizing principle of human socio-emotional experience, based on the discovery that these dimensions structure the semantic relations of linguistic concepts across languages and cultures~\cite{Osgood1975}. They also emerged from statistical analyses of the co-occurrence of a large variety of physiological, facial, gestural, and cognitive features of emotional experience~\cite{Fontaine2007}, and relate to the universal dimensionality of personality, and social cognition~\cite{Scholl2013}. 

Combined, Evaluation, Potency, and Activity form the EPA tri\-ple, which encapsulates a sentiment in ACT. Sociologists have gathered such values for many common words by conducting large, cross-cultural studies~\cite{heise2010surveying,osgood1975cross}. As such, they represent the fundamental (or out-of-context) sentiments held by the particular group surveyed.  Affect control theorists have compiled lexicons of a few thousand  words along with average EPA ratings obtained from survey participants who are knowledgeable about their culture~\cite{heise2010surveying}. For example, most English speakers agree that professors are about as nice as students (E), more powerful (P) and less active (A). The corresponding EPAs (called {\em identity sentiments} or {\em identities} when referring to social roles of people) are $[1.7,1.8,0.5]$ for professor and $[1.8,0.7,1.2]$ for student.\footnote{\label{foot:actlex} All EPA labels and values in this paper are taken from the Indiana 2002-2004 ACT lexicon~\cite{heise2010surveying} unless otherwise noted.}


An individual's feelings about an entity (person/thing/ac\-tion) may change temporarily when that entity is observed in some context~\cite{heise1977social}. Consider for example a professor (actor) who yells (behaviour) at a student (object). Most would agree that this professor appears considerably less nice (E), a bit less potent (P), and certainly more active (A) than the cultural average of a professor. To formalize this idea, ACT defines transient sentiments (called transient {\em impressions}) to be EPA values that have been altered by context. In this case, context entails an Actor-Behaviour-Object (A-B-O) interaction, in which an Actor (A) performs a Behaviour (B) on an Object (O). The transient shifts in affective meaning caused by specific events are described using models of the form $\trb=\ACTM\Gop(\fub)$, $\trb$ is the transient impression of the A-B-O triple (a nine-dimensional vector), $\fub$ is the fundamental sentiment (also a 9D vector), $\ACTM$ is a globally defined matrix of statistically estimated prediction coefficients from empirical impression-formation studies and $\Gop$ is a vector of polynomial features in $\fub$ that are hand-crafted to reflect important relationships between the various sentiments. The squared Euclidean distance between fundamental and transient sentiments increases with the unexpectedness of the interaction, and is called the deflection. It is from this quantity that ACT draws its predictive power as a model of human behaviour. It states that people naturally act in a way that minimizes the deflection they create \cite{heise1977social}. That is, one's default action is that which aligns best with society's expectations. Given an Actor and an Object, it is therefore possible to predict a Behaviour by choosing for it an EPA profile that minimizes the deflection.  Certainly, and agent may choose to act in a different way, but then must incur the penalties of deflection. This may entail feelings of being disingenuous to one's true self (discomfort, guilt, etc.), as well as receiving a re-identification in the eyes of any observers or interactants. Such re-identifications can significantly alter the subsequent course of the interaction.

ACT has been shown to be highly accurate in explaining verbal behaviours of mock leaders in a computer-simulated business~\cite{Schroeder2009}, non-verbal displays in dyadic interactions~\cite{Schroeder2013}, and group dynamics~\cite{Heise2013}, among others~\cite{MacKinnon2014}. The case of group dynamics is of particular interest for our work, as it presents a sophisticated model of interactant selection based on perceived deflections and feelings of inauthenticity~\cite{Heise2013}.  This ACT-based model was applied to simulated jury deliberation data in~\cite{Heise2013}, and found to replicate important properties of the human data. As such, it could form a future extension of the work we present here.

BayesACT is a recently proposed generalisation of ACT, adding three new elements~\cite{hoey2013bayesian}. First, sentiments are viewed as probability distributions over latent variables rather than points in the EPA space, allowing for multimodal, uncertain and dynamic sentiments to be modeled and learned.  One frequently encounters situations where multiple identities may be mixed, and multiple sentiments may occur simultaneously~\cite{SmithLovin2007}. Second, affective interactions are augmented with {\em denotative} state (e.g. the usual state space considered in typical AI applications). Third, an explicit reward function allows for goals that go beyond simple deflection minimization.  The BayesACT model is formulated as a partially observable Markov decision process (POMDP), and a Monte-Carlo tree search method based on the POMCP algorithm~\cite{silver2010monte} is used to generate a policy of action. In this method, nodes in a decision tree are expanded with probability inversely proportional to the deflection their respective action creates. That is to say, a desire to minimize the deflection guides the direction in which the tree expands, but the actual action taken is ultimately decided by maximization of the reward received. This is accomplished by repeated sampling of the state space distribution for as much time as is allowed, where a longer running time results in a more densely filled tree, and therefore a more "scheming" agent.  We refer the reader to the complete description of the model in~\cite{HoeyBACT15,SchroederHoeyRogers2016}. 

When applied to the prisoner's dilemma game, we use the method described by Asghar and Hoey~\cite{asghar2015intelligent}, in which the two {\em identities} (the players in the game) are set to a be probabilistic random mixture of \elab{friend} \epa{2.8}{1.9}{1.4} and \elab{scrooge} (EPA:$\{\mi 2.2,\mi 0.2,\mi 0.5\}$). Thus, if an agent thought of himself as a \elab{friend} and knew the other agent to be a \elab{friend}, the deflection minimizing action would likely be something good (high E), which is closer (affectively) to the cooperation action (e.g. {\em collaborate with} \epa{1.44}{1.11}{0.61}), rather than the defection action (e.g. {\em abandon} \epa{\mi 2.3}{\mi 0.5}{\mi 0.8}). Using this method, we can compute that a \elab{friend} would still collaborate with a \elab{scrooge} (in an attempt to \elab{reform} the scrooge), while a \elab{scrooge} would abandon a \elab{friend} (\elab{look away from} in shame), and two scrooges would defect.

Another way of assigning identities for agents is to measure human identity sentiments while playing the PD game with other humans.  For this, we used the data from the work of Jung {\em et al.}~\cite{jung2016grounding}, in which the affective sentiments of PD game players were measured via surveys. These results provide actual initial distributions rather than simple points in EPA space (as in the {\em friend/scrooge} case) for the agent's self-identity and its perception of its opponent's identity. Additionally, sentiments were similarly collected for each of the two actions, cooperate and defect. Though BayesACT does not currently support distributions for these, their averages provide points in EPA space that are tuned specifically for the purposes of the Prisoner's Dilemma. Further discussion of these values may be found in the ``Experimental Parameters'' section.

Our work differs from that in Asghar and Hoey~\cite{asghar2015intelligent}, as we are interested in the networked version of the game, and (importantly), we conduct an explicit quantitative comparison between the results of BayesACT and those of real human study participants. Additionally, as BayesACT is designed to keep a single EPA distribution for its self-identity and another such distribution for its opponent, it is not immediately equipped to handle multiple opponents (and hence, a networked setting). We have therefore modelled this multiple-opponent case as a single game against an aggregate opponent, whose actions have an EPA equal to the average EPA of those from all individual opponents. This approach allows each BayesACT agent to have a consistent self-sentiment and permits all interactions to occur simultaneously.

\subsection{The Prisoner's Dilemma}

Study of the Iterated Prisoner's Dilemma dates back to 1952, with the experiments of Merrill Flood and Melvin Dresher~\cite{flood1958some}. In this work and others that followed~\cite{andreoni1993rational, deutsch1958trust}, it is observed that human participants do not play the game rationally (in the game theoretic sense). This is easiest to see in the single-shot version of the game, which has defection as a dominant strategy, yet produces high rates of cooperation from human players~\cite{deutsch1958trust}.

It can also be shown that pure defection is a dominant strategy for any version of the Iterated Prisoner's Dilemma for which the number of rounds is known by the players~\cite{kreps1982rational}. However, studies have shown that this is not the way that humans behave~\cite{andreoni1993rational,deutsch1958trust,flood1958some}. Clearly, there are more factors at play than reward maximization.

Explanations for this cooperative behaviour have been linked to the phenomenon of altruism in nature~\cite{nowak2006five}, which may give an evolutionary benefit through factors like kinship, direct reciprocity, or indirect reciprocity via reputation. Trivers~\cite{trivers1971evolution} argues that such altruism would almost certainly have been an advantageous trait for early groups of humans to have, citing cases that come at a small cost to the giver, but result in a large benefit for the receiver. Examples include sharing food and tools, helping the wounded, sick, or very young, and sharing knowledge. Further, small, stable groups would have provided ample opportunity for acts of altruism to be applied to kin, or to be reciprocated by the receiver in the future. As a means of encouraging such acts, Trivers proposes the development (or at least co-option) of emotion. Sympathy is an impetus to help those in need. Gratitude promotes returning the favour. Guilt dissuades from cheating others in the group.

This is all to say that a person playing the Prisoner's Dilemma might choose not to defect out of guilt, which, to that person, is simply an immediate emotional response. However, that guilt exists as a means of promoting altruism through a long evolutionary process. By application of ACT, we can now predict when such emotional intervention will occur.

\subsection{Networked Prisoner's Dilemma Games \label{sec:networkPD}}

When generalizing the Prisoner's Dilemma to include more than two players (ie. the N-Person [Iterated] Prisoner's Dilemma), we must choose how to expand the reward matrix to accommodate the additional agents. In this work, we employ the Broadcast model, in which an agent chooses to either cooperate or defect, and this choice is applied to all partners (neighbours) according to a matrix in the form described by Table \ref{tab:PDGame}. This framework is desirable because it provides an opportunity to compare our results to those of studies conducted with human participants, usually (but not always) arranged on a grid \cite{grujic2010social,grujic2012consistent,rand2011dynamic}.

Gruji\'c et al. \cite{grujic2014comparative} examine several such studies and make five observations regarding general human player behaviour:

\begin{para_packed_enum}
    \item \textbf{Network Invariance} - Once the number of neighbours has exceeded two, the average degree and structure of the network cease to significantly affect cooperation rates.
    \item \textbf{Declining Cooperation Over Time} - The global cooperation rate begins high, but asymptotically declines to a near constant, but non-zero value.
    \item \textbf{Anti-Correlation of Earnings and Cooperation} - On average, cooperation yields a lower reward than defection. This comes as a result of defectors performing better in a mixed environment.
    \item \textbf{Moody Conditional Cooperation (MCC)} - The largest fraction of players exhibit behaviour consistent with Moody Conditional Cooperation, a strategy in which the probability of an agent cooperating increases with the number of cooperating neighbours, but also has a hysteresis. That is, an agent who cooperated last turn is more likely to cooperate again than one who defected last turn, assuming the two have the same number of cooperating neighbours.
    \item \textbf{Player Type Stratification} - The remaining players are stratified into the other four groups identified by Gruji\'c et al. \cite{grujic2010social} (Pure Cooperators, Mostly Cooperators, Mostly Defectors, and Pure Defectors).
\end{para_packed_enum}

Our goal is to build automated strategies to replicate these findings about human play.  However, simple deterministic agents, like tit-for-tat (which copies its partner's last move), while highly effective in non-networked PD games, turn out to be uninteresting in the networked setting. Plain tit-for-tat, in particular, produces a network of either full cooperation or full defection in just a few iterations.  However, more complex imitation-based models replicate at least the MCC property. For example, in the model we compare against in this paper~\cite{vilone2014social}, every simulated agent plays by the following strategy:

\begin{para_packed_item}
    \item With probability $q$, imitate the action of a randomly chosen neighbour.
    \item With probability $(1-q)$, perform a strategic imitation. (Several different methods were tried, but the main one presented in~\cite{vilone2014social} is Unconditional Imitation, in which the single best scoring neighbour, including self, is imitated.)
\end{para_packed_item}


\section{Description of the Experiment}

For each test, 169 bots of one type (ie. BayesACT or imitation), where 169 was chosen to correspond with a 13 by 13 grid, were arranged on a static network to play the Iterated Prisoner's Dilemma with their neighbours. These games each lasted for 60 individual rounds (or iterations), a number comparable to those of the largest human studies \cite{grujic2014comparative}. For each setting of our test parameters (to be described shortly), 20 independent games were played, resulting in approximately 3000 total simulations~\footnote{(17 agents)*(9 matrix/network combinations)*(20 simulations each) = 3060 simulations}.

Each round, agents chose between cooperation and defection and relayed that choice to each of their partners (where partners are defined to be neighbours on one of the networks described below), thus adhering to the Broadcast model. Scores were then determined by summing the rewards earned in each of the resulting one-on-one games. Partners were selected on the basis of static network connections determined before starting the game.

\subsection{Experimental Parameters}

\subsubsection{Network Types}

We selected two classes of networks on the basis of their use in either human studies \cite{grujic2014comparative} or the work of Vilone et al. \cite{vilone2014social} on imitation-based agents:
\begin{para_packed_item}
    \item Grid - Agents are arranged on a square grid and interact with their nearest neighbours. These neighbours may belong to either a Moore (four neighbours, cardinal directions only) or von Neumann (eight neighbours, diagonals included) neighbourhood.
    \item Erd\"os-R\'enyi (ER) - Given a desired average node degree, all networks with that property are equally likely to be chosen. ie. Edges are chosen randomly with uniform probability until some threshold is reached.
\end{para_packed_item}

Given that these network types (which, of course, are not the only possible valid networks) can be varied in infinite ways (for instance, by density), we had a free choice over which would actually be incorporated into this experiment. The selections we made are as follows:
\begin{para_packed_enum}
    \item a grid layout with a Moore neighbourhood (Grid),
    \item an ER network with average degree 5.14 (ER5),
    \item an ER network with average degree 8.48 (ER8).
\end{para_packed_enum}
\noindent where the particular values of 5.14 and 8.48 were chosen to mirror those employed by Vilone et al.~\cite{vilone2014social} It is additionally convenient that our first and third networks had similar density, differing primarily in the regularity of their connections.

\subsubsection{Reward Matrices}

Despite (or perhaps because of) the enormous number of Prisoner's Dilemma studies, there has been little agreement on reward values outside the basic inequality given by Equation \ref{eq:PDInequality}. They therefore present another free choice. In this experiment, we use the matrices given by Table \ref{tab:rewardMatrices}. The first matrix (M1) satisfies the Prisoner's Dilemma inequality with the smallest possible non-negative integers, the second (M2) is one setting employed by Vilone et al.~\cite{vilone2014social}, and the third (M3) is the one used by Asghar and Hoey \cite{asghar2015intelligent}. To a large extent, it is not any particular set of matrix values that is important, but rather, the difference between them. In particular, defection is far more beneficial on M1 than on M3, which creates two very different situations for bots to handle.
In fact, Vilone et al.~\cite{vilone2014social} tested a continuum of matrices like M2 by varying $T$ (The `temptation' to defect, as in Table \ref{tab:PDGame}). Our M2 uses a value of $T$ that saw a large variation in behaviour depending on the value of their randomization parameter, $q$. M2 does not adhere to the strict inequality of Equation \ref{eq:PDInequality}, but nevertheless preserves the central difficulty of the Prisoner's Dilemma (namely, that a rational agent defects while the global optimum is for cooperation).

\begin{table}[tb]
    \centering{
        \begin{tabular}{cc}
            M1
            \begin{tabular}{c|c|c}
                & C   & D   \\
                \hline
                C & 2,2 & 0,3 \\
                \hline
                D & 3,0 & 1,1 \\
            \end{tabular}
            &
            M2
            \begin{tabular}{c|D{,}{,}{3.1}|D{,}{,}{1.3}}
                & C   & D   \\
                \hline
                C & 1,1 & 0,1.4 \\
                \hline
                D & 1.4,0 & 0,0 \\
            \end{tabular}
            \\
        \end{tabular}
        \newline
        M3
        \begin{tabular}{c|D{,}{,}{2.2}|D{,}{,}{1.2}}
            & C   & D   \\
            
            \hline
            C & 10,10 & 0,11 \\
            \hline
            D & 11,0 & 1,1 \\
        \end{tabular}
        
        \caption{\label{tab:rewardMatrices} Reward matrices chosen for our experiments.}
    }
\end{table}

\subsubsection{Other Parameters}

Additionally, each of the two bots tested had their own unique parameters. In the case of BayesACT, we chose to vary the initial EPA distribution between the original set as presented by Hoey et al. \cite{hoey2013bayesian} and one gleaned from a human study by Jung et al. \cite{jung2016grounding}. In the former, the cooperate and defect actions were assigned the EPAs of ``flatter'' \epa{2.1}{1.5}{0.8} and ``abandon'' \epa{-2.3}{-0.5}{-0.8} respectively, while in the latter, they were found to be \epa{1.4}{0.1}{0.2} and \epa{-0.7}{0.9}{0.7} by surveying human participants. We also applied several different computation time limits (0, 1, and 10 seconds) to BayesACT's POMCP search. When a particular setting of these parameters must be identified, we use the shorthand BACT[X][Y], where X is one of D or S (denoting default or study EPA settings) and Y is one of 0, 1, or 10 (denoting the POMCP timeout). 

For the imitation-based bots, we varied $q$, the probability of randomly selecting any neighbour instead of the highest scorer, from 0\% to 100\% in 10\% intervals. We identify this via the shorthand IM[X], where $X \in [0,100]$ is the value of $q$. 

\section{Results}

We now examine the results of these experiments with respect to the five observations of Gruji\'c et al.~\cite{grujic2014comparative}. For nominal variables, we have applied the G-test to show independence (or lack thereof) of two distributions.  The G-test is the likelihood ratio $G = 2 \sum_{i}O_i log(O_i/E_i)$, where $O_i$ is the observed count and $E_i$ is the expected count \cite{mcdonald2014multiple}. This can be converted to a p-value, for which we consider a value of $0.05$ to constitute strong evidence that the distributions in question are different.
For continuous variables, we used ANOVA, which produces a similar p-value.

\subsection{Network Invariance}
For all parameter settings of the BayesACT agents, we do not find evidence that network structure impacts agent behaviour. This is demonstrated by the consistently high p-values obtained when performing a G-test of cooperation rate per round across the 3 network types for a particular reward matrix, indicating that the three distributions are not statistically discernible. In particular, for BACTD agents, 96.1\% of rounds (each corresponding to one table row) have $p>0.05$, while for BACTS agents, this is true of 90.7\% of rows, where 100\% corresponds to 540 total table rows (60 rounds/game X 3 reward matrices X 3 BayesACT parameter settings). As an example, Table \ref{tab:bactNetInvar} gives these data for one BayesACT setting, and for matrix M1. Notice that the cooperation rates for each of the various networks within a row are within just a few percent of each other. A G-test shows that these small differences are, in fact, not statistically significant, indicating network invariance.

\begin{table}[tb]
    \centering{
        \begin{tabular}{c|N|N|N|N|N}
            Round &  \mcb {Grid}  & \mcb {ER5}    & \mcb {ER8}    & \mcb {G-test} & \mcnb {G-test}  \\
            & \mcb {C \%}   & \mcb {C \%}    & \mcb {C \%}    & \mcb {stat.}     & \mcnb {p-val.} \\
            \hline
            \rowcolor{LightCyan}
            R0 & 32.81 & 33.05 & 33.46 & 0.33 & 0.85 \\
            \rowcolor{LightCyan}
            R1 & 33.67 & 33.76 & 33.64 & 0.01 & 0.99 \\
            \rowcolor{LightCyan}
            R2 & 31.12 & 32.51 & 32.87 & 2.64 & 0.27 \\
            \rowcolor{LightCyan}
            R3 & 30.50 & 30.00 & 31.54 & 1.95 & 0.38 \\
            \rowcolor{LightCyan}
            R4 & 30.30 & 30.15 & 30.24 & 0.02 & 0.99 \\
            ...& ...  & ...  & ...  & ...  & ...   \\
            \rowcolor{LightCyan}
            R57 & 29.23 & 28.17 & 28.79 & 0.95 & 0.62 \\
            \rowcolor{LightCyan}
            R58 & 27.43 & 27.31 & 28.14 & 0.68 & 0.71 \\
            \rowcolor{LightCyan}
            R59 & 28.22 & 28.02 & 28.99 & 0.88 & 0.64 \\
        \end{tabular}
    }
    \caption{\label{tab:bactNetInvar} Comparing cooperation rates by round across our three network types for a BayesACT agent using the default EPA distribution, a timeout of 1 second, and the first of our three reward matrices (M1). Rows are shaded for p-values $> 0.05$, indicating that the distributions for the three network type are not statistically discernible.}
\end{table}

Imitation-based agents, on the other hand, do not seem to consistently exhibit this invariance. For IM agents, only 5.6\% of rounds have $p > 0.05$, where 100\% corresponds to 1980 total table rows (60 rounds/game X 3 reward matrices X 11 imitator parameter settings). We find that there is a strong tendency for these agents to gradually move towards full defection, but they frequently do so at different rates, depending on the network (with the cooperation rate of the comparatively low density ER5 network frequently decaying the slowest~\footnote{In fact, if we consider only the Grid and ER8 networks, which have very similar average degree, the number of rows with $p > 0.05$ increases significantly to 27.7\%, indicating that density may have a large effect on the behaviour of imitator agents}). Further, when full defection has been reached for one network, any deviation in the others becomes statistically significant. This results in low G-test p-values when observing cooperation rates on a per-round basis, indicating that behaviour is indeed affected by network structure.  An example of this behaviour is given by Table \ref{tab:imitNetInvar}, where we see large differences in cooperation rates within a row, particularly between ER5 and the two other networks. Very low G-test p-values ($p << 0.05$) indicate that these differences are statistically significant and that network invariance is therefore not observed.

\begin{table}[tb]
    \centering{
        \begin{tabular}{c|N|N|N|N|N}
            Round    &  \mcb{Grid}  &  \mcb{ER5}    &  \mcb{ER8}    & \mcb{G-test}    & \mcnb{G-test}  \\
            &  \mcb{C \%}   &  \mcb{C \%}    &  \mcb{C \%}    & \mcb{stat.}     & \mcnb{p-val.} \\
            \hline
            \rowcolor{LightCyan}
            R0 & 51.36 & 49.67 & 48.91 & 4.27 & 0.12 \\
            R1 & 25.15 & 30.83 & 26.09 & 31.18 & <0.001 \\
            R2 & 12.69 & 17.10 & 12.96 & 33.01 & <0.001 \\
            R3 & 6.27 & 8.82 & 5.98 & 24.46 & <0.001 \\
            R4 & 2.75 & 4.76 & 3.11 & 22.00 & <0.001 \\
            ...& ...  & ...  & ...  & ...  & ...   \\
            R57 & 0.00 & 0.24 & 0.03 & 13.51 & 0.001 \\
            R58 & 0.00 & 0.24 & 0.03 & 13.51 & 0.001 \\
            R59 & 0.00 & 0.24 & 0.03 & 13.51 & 0.001 \\
        \end{tabular}
    }
    \caption{\label{tab:imitNetInvar} Comparing cooperation rates by round across our three network types for an imitation agent using $q=0.5$ and the first of our three reward matrices (M1). Rows are shaded for p-values $> 0.05$, indicating that the distributions for the three network type are not statistically discernible.}
\end{table}

\subsection{Cooperation Rate Over Time}

In human studies, the global cooperation rate has been observed to drop from 55\%-70\% to less than 20\%-40\% after around 20 rounds of play, after which it remains approximately constant \cite{grujic2014comparative}. We do not in general observe this behaviour in our BayesACT agents. While some settings of BayesACT do demonstrate a reduction in global cooperation rate over time (ie. round number), this difference is typically less than 5\%

The imitation-based agents, on the other hand, tended to display one of two extreme behaviours: either the cooperation rate decayed to zero, or it ballooned to some constant greater that its starting value. In some cases, individual simulations of the same parameter settings were split between these final states, though a descent into full defection was generally the more common of two. Such behaviour is also decidedly non-human.

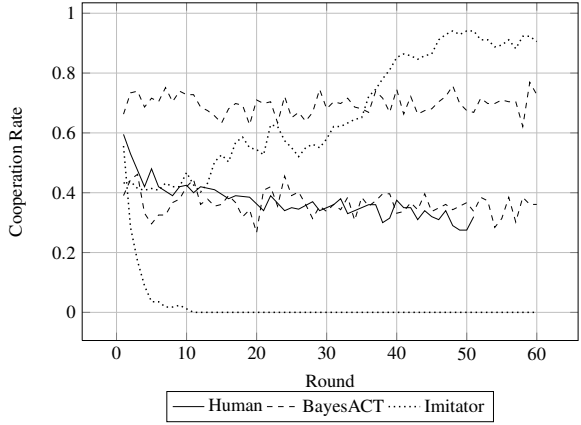
\begin{figure}[tb]
    \centering{
        
        \begin{tikzpicture}[scale=0.7]
        \begin{axis}[
        height=8cm,
        width=11cm,
        grid=major,
        xlabel=Round,
        ylabel=Cooperation Rate,
        legend style={at={(0.5,-0.15)},
            anchor=north,legend columns=3}
        ]
        
        
        \addplot+[mark=none,black] table{
            1 0.595
            2 0.53
            4 0.42
            5 0.48
            6 0.42
            7 0.405
            8 0.39
            9 0.42
            10 0.425
            11 0.4
            12 0.42
            14 0.41
            16 0.38
            17 0.39
            19 0.385
            21 0.34
            22 0.39
            24 0.34
            25 0.35
            26 0.345
            28 0.37
            29 0.34
            31 0.36
            32 0.38
            33 0.33
            34 0.34
            35 0.35
            36 0.36
            37 0.36
            38 0.3
            39 0.315
            40 0.375
            41 0.35
            42 0.35
            43 0.31
            44 0.34
            45 0.32
            46 0.31
            47 0.34
            48 0.29
            49 0.275
            50 0.275
            51 0.32
        };
        \addlegendentry{Human}
        
        \addplot+[dashed,mark=none,black] table {
            1 0.3905
            2 0.4438
            3 0.4615
            4 0.3314
            5 0.2959
            6 0.3254
            7 0.3254
            8 0.3669
            9 0.3787
            10 0.4201
            11 0.4438
            12 0.3609
            13 0.3787
            14 0.3550
            15 0.3609
            16 0.3905
            17 0.3669
            18 0.3195
            19 0.3432
            20 0.2663
            21 0.4083
            22 0.4201
            23 0.3550
            24 0.4556
            25 0.3905
            26 0.4024
            27 0.3609
            28 0.3136
            29 0.3550
            30 0.3373
            31 0.3609
            32 0.3432
            33 0.3846
            34 0.3077
            35 0.3846
            36 0.3550
            37 0.3728
            38 0.3964
            39 0.3964
            40 0.3314
            41 0.3373
            42 0.3669
            43 0.3432
            44 0.3964
            45 0.3373
            46 0.3491
            47 0.3609
            48 0.3432
            49 0.3550
            50 0.3669
            51 0.3373
            52 0.3846
            53 0.3728
            54 0.2840
            55 0.3136
            56 0.3846
            57 0.3018
            58 0.3846
            59 0.3609
            60 0.3609
            
        };
        \addlegendentry{BayesACT}
        
        \addplot+[dotted, thick, mark=none,black] table {
            1 0.5562
            2 0.2840
            3 0.1716
            4 0.0888
            5 0.0355
            6 0.0355
            7 0.0178
            8 0.0178
            9 0.0237
            10 0.0118
            11 0.0000
            12 0.0000
            13 0.0000
            14 0.0000
            15 0.0000
            16 0.0000
            17 0.0000
            18 0.0000
            19 0.0000
            20 0.0000
            21 0.0000
            22 0.0000
            23 0.0000
            24 0.0000
            25 0.0000
            26 0.0000
            27 0.0000
            28 0.0000
            29 0.0000
            30 0.0000
            31 0.0000
            32 0.0000
            33 0.0000
            34 0.0000
            35 0.0000
            36 0.0000
            37 0.0000
            38 0.0000
            39 0.0000
            40 0.0000
            41 0.0000
            42 0.0000
            43 0.0000
            44 0.0000
            45 0.0000
            46 0.0000
            47 0.0000
            48 0.0000
            49 0.0000
            50 0.0000
            51 0.0000
            52 0.0000
            53 0.0000
            54 0.0000
            55 0.0000
            56 0.0000
            57 0.0000
            58 0.0000
            59 0.0000
            60 0.0000
            
        };
        \addlegendentry{Imitator}
        
        \addplot+[dashed,mark=none,black] table {
            1 0.6627
            2 0.7337
            3 0.7396
            4 0.6864
            5 0.7160
            6 0.7041
            7 0.7515
            8 0.7041
            9 0.7396
            10 0.7278
            11 0.7278
            12 0.6864
            13 0.6746
            14 0.6568
            15 0.6331
            16 0.6805
            17 0.6982
            18 0.6923
            19 0.6272
            20 0.7101
            21 0.6982
            22 0.7041
            23 0.6331
            24 0.7219
            25 0.6509
            26 0.6686
            27 0.6391
            28 0.6686
            29 0.7456
            30 0.6805
            31 0.7041
            32 0.6982
            33 0.7160
            34 0.6864
            35 0.6805
            36 0.6686
            37 0.7396
            38 0.7160
            39 0.6686
            40 0.7456
            41 0.6627
            42 0.7219
            43 0.6627
            44 0.6746
            45 0.6805
            46 0.7041
            47 0.7219
            48 0.7574
            49 0.6982
            50 0.6746
            51 0.6686
            52 0.7160
            53 0.6982
            54 0.6982
            55 0.7101
            56 0.7041
            57 0.7041
            58 0.6213
            59 0.7692
            60 0.7278
            
        };
        
        \addplot+[dotted, thick, mark=none,black] table {
            1 0.4320
            2 0.4379
            3 0.4142
            4 0.4083
            5 0.4142
            6 0.4083
            7 0.4320
            8 0.4201
            9 0.4201
            10 0.4675
            11 0.4260
            12 0.4024
            13 0.4320
            14 0.5030
            15 0.5266
            16 0.5030
            17 0.5680
            18 0.5858
            19 0.5503
            20 0.5444
            21 0.5266
            22 0.6272
            23 0.6213
            24 0.5740
            25 0.5503
            26 0.5207
            27 0.5503
            28 0.5621
            29 0.5503
            30 0.5799
            31 0.6213
            32 0.6213
            33 0.6331
            34 0.6450
            35 0.6509
            36 0.7219
            37 0.7456
            38 0.7811
            39 0.8107
            40 0.8521
            41 0.8639
            42 0.8580
            43 0.8462
            44 0.8580
            45 0.8639
            46 0.9112
            47 0.9290
            48 0.9408
            49 0.9290
            50 0.9408
            51 0.9408
            52 0.9112
            53 0.9112
            54 0.8876
            55 0.8935
            56 0.9112
            57 0.8817
            58 0.9231
            59 0.9231
            60 0.9053
        };
        
        \end{axis}
        \end{tikzpicture}
        
        \caption{\label{fig:coopRates} Cooperation rates by round for human players~\protect\cite{gracia2012heterogeneous} (which had matrix parameters T=10, R=7, P=0, S=0, and used a grid layout with a von Neumann neighbourhood)~\protect\cite{gracia2012heterogeneous}, and selected simulations for BayesACT (bottom: BACTS1/M1/ER5, top: BACTS0/M1/Grid) and imitator (bottom: IM50/M2/Grid, top: IM100/M3/Grid) agents.}
    }
\end{figure}

Figure~\ref{fig:coopRates} shows an example of the cooperation rate over time for human players (taken from~\cite{gracia2012heterogeneous}) and for two simulations each of imitation-based and BayesACT agents. The slow decline but stabilisation of human players can be seen, while the IM simulations were selected to demonstrate the two behaviours described above (i.e. a quick descent into full defection or a ballooning cooperation rate). On the other hand, BayesAct simulations produce more stable cooperation rates, some of which are relatively close to those of humans (after stabilization), and some of which are not (as our two selected simulations demonstrate). In any case, we do not generalize the initial rapid drop in cooperation rate characteristic of human players.

\begin{figure}
    \includegraphics[width=\columnwidth]{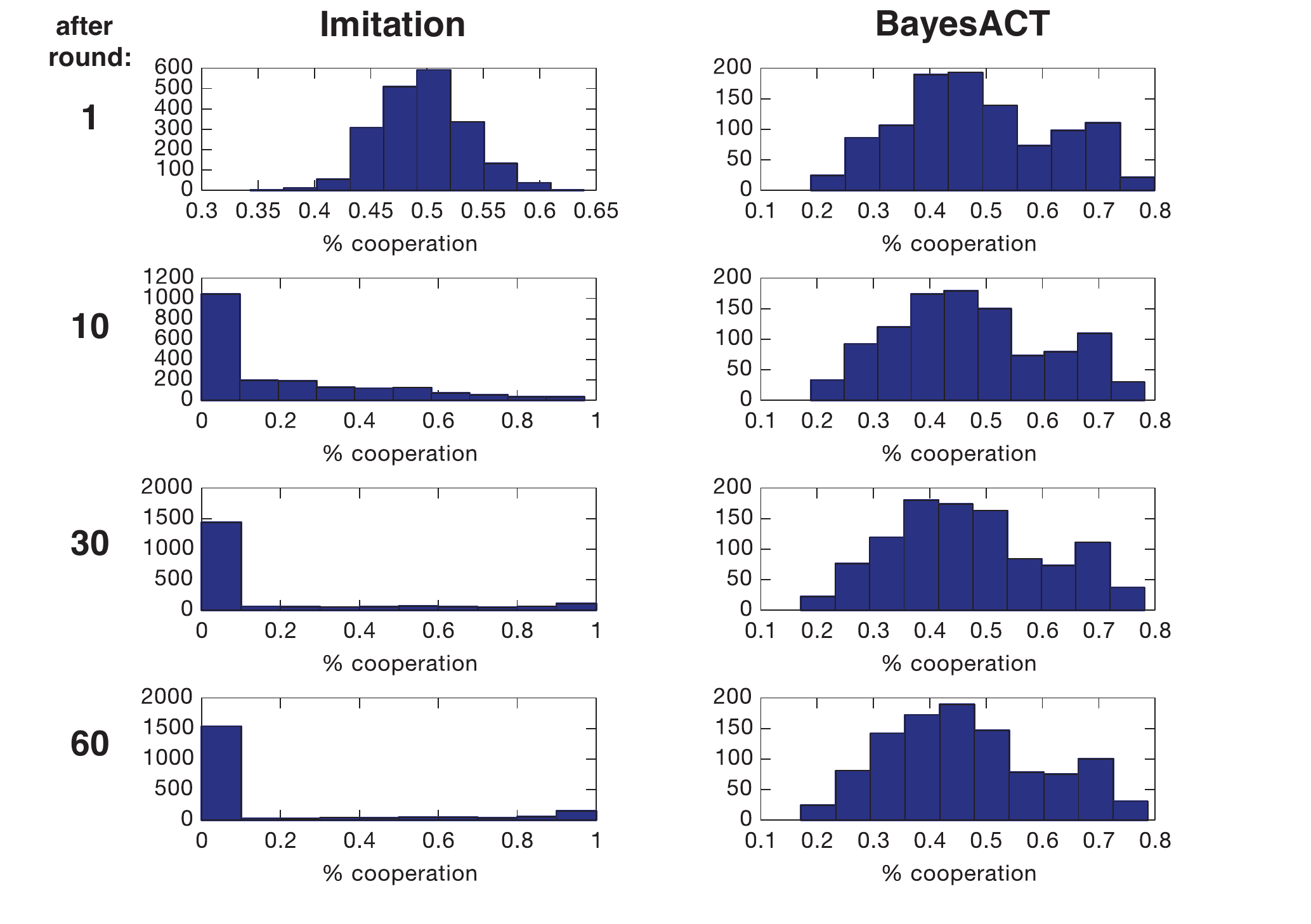}
    \caption{\label{fig:bactHist} Histograms of the number of simulations at various cooperation rates for imitator agents (left) and BayesACT agents (right) at rounds 0, 10, 30, and 60 (top to bottom). BayesACT agents tend to remain fairly constant in their behaviour, while imitator agents have a strong tendency towards full defection (and to a lesser extent, full cooperation).}
\end{figure}

Distributions over cooperation rates after $1,10,30$ and $60$ rounds are shown in Figure~\ref{fig:bactHist}. Again, we see the imitation agents rapidly have almost all converged to all defect, while BayesACT agents remain more stable in the middle (just less than 50\% cooperation. 

\subsection{Anti-Correlation of Earnings and Cooperation}

Before calculating correlation coefficients, it is worth examining the relative scores of cooperators and defectors, which ought to be higher for the defectors. Across all parameter settings, BayesACT agents score lower when cooperating than when defecting (i.e. test passed for 100\% of both BACTD and BACTS agents). While the imitator agents do display this property for Matrix 1 (for 100\% of settings), they do not generally do so for Matrices 2 and 3 (6.7\% and 0\% of settings, where the only successful agents were the purely random imitators, IM100). It is important to note that reward matrix values are parameters of the simulation, not of the agents. That is to say, a successful agent must have at least some parameter setting that can cope with any possible reward matrix. A failure of the imitator agents for M3 (and M2 as well, if we exclude IM100) indicates a failure overall.

Table \ref{tab:anticor} gives an example of these results, where we see that BayesACT simulations produce higher average scores for defectors (the desired result), while IM simulations show just the opposite. In all cases, there was a statistically significant difference between the scores of cooperators and defectors, as calculated by ANOVA.

\begin{table}[tb]
    \centering{
        \begin{tabular}{c|N|N|c|N}
            Agent &  \mcb{Avg. C}  &  \mcb{Avg. D} & \mcb{ANOVA} & \mcnb{ANOVA} \\
            &  \mcb{Score}   &  \mcb{Score}  & \mcb{stat.}     & \mcnb{p-val.} \\
            \hline
            \rowcolor{LightCyan}
            BACTD0 & 25.72 & 30.54 & \num{4.01E+3} & <0.001 \\
            \rowcolor{LightCyan}
            BACTD1 & 23.67 & 28.64 & \num{4.78E+3} & <0.001 \\
            \rowcolor{LightCyan}
            BACTD10 & 19.61 & 24.97 & \num{6.27E+3} & <0.001 \\
            \rowcolor{LightCyan}
            BACTS0 & 35.69 & 40.82 & \num{3.09E+3} & <0.001 \\
            \rowcolor{LightCyan}
            BACTS1 & 32.52 & 37.48 & \num{3.40E+3} & <0.001 \\
            \rowcolor{LightCyan}
            BACTS10 & 25.33 & 30.20 & \num{4.35E+3} & <0.001 \\
            \hline
            IM0 & 45.19 & 22.51 & \num{4.05E+4} & <0.001 \\
            IM10 & 47.41 & 21.81 & \num{4.20E+4} & <0.001 \\
            IM20 & 46.11 & 19.14 & \num{6.14E+4} & <0.001 \\
            IM30 & 42.35 & 21.69 & \num{4.28E+4} & <0.001 \\
            IM40 & 35.92 & 14.03 & \num{6.59E+4} & <0.001 \\
            IM50 & 26.89 & 11.96 & \num{3.49E+4} & <0.001 \\
            IM60 & 19.80 & 8.51 & \num{2.08E+4} & <0.001 \\
            IM70 & 19.58 & 8.87 & \num{1.87E+4} & <0.001 \\
            IM80 & 20.41 & 9.06 & \num{2.06E+4} & <0.001 \\
            IM90 & 22.97 & 12.69 & \num{1.58E+4} & <0.001 \\
            IM100 & 25.12 & 22.57 & \num{9.57E+2} & <0.001 \\
        \end{tabular}
    }
    \caption{\label{tab:anticor} Comparing scores earned by cooperators and defectors among different agent types. Shaded rows indicate that the average score for defection is higher than that of cooperation, as per Requirement 3, and that the difference is statistically significant ($p<0.05$). This table presents results for network ER5 and Reward M3.}
\end{table}

We see similar results when calculating the Pearson Correlation between the cooperation rates of individual agents and their scores (normalized by the average score of all agents in their simulation). To consider Requirement 3 satisfied, we must see a negative correlation coefficient that is statistically significant ($p<0.05$). By this metric, we find that 100\% of BACTD settings and 74.1\% of BACTS settings (100\% for M1, 88.9\% for M2, 33.3\% for M3) display the desired anti-correlation. Given that even the best BACTS settings (BACTS0 or BACTS10) only achieved 77.8\% requirement satisfaction, they may be considered less successful than their BACTD counterparts. However, this difference is small compared to the imitator agents, which displayed cooperation-score anti-correlation in only 30.3\% of settings (81.8\% for M1, 9.1\% for M2, 0\% for M3), with the best setting, IM10, succeeding in 44.4\% of matrix/network combinations. By both of our metrics, then, BayesACT satisfies Requirement 3 more adequately than the imitator bots.

Table \ref{tab:corCoef} gives an example of these results, where we see that all BACTD agents achieve the desired negative Pearson coefficient with a statistically significant p-value, which is not the case for either the BACTS or IM agents.

\begin{table}[tb]
    \centering{
        \begin{tabular}{c|N|N|N|N}
            Agent &  \mcb{Avg.}  &  \mcb{Avg.} & \mcb{Pearson} & \mcnb{Pearson} \\
            &  \mcb{C \% }   &  \mcb{Score}  & \mcb{coef.}     & \mcnb{p-val.} \\
            \hline
            \rowcolor{LightCyan}
            BACTD0 & 49.79 & 28.13 & -0.10 & <0.001\\
            \rowcolor{LightCyan}
            BACTD1 & 45.93 & 26.36 & -0.12 & <0.001\\
            \rowcolor{LightCyan}
            BACTD10 & 38.60 & 22.90 & -0.16 & <0.001\\
            BACTS0 & 69.49 & 37.26 & -0.02 & 0.212\\
            BACTS1 & 63.14 & 34.35 & 0.00 & 0.949\\
            BACTS10 & 48.99 & 27.81 & 0.00 & 0.857\\
            \hline
            IM0 & 74.06 & 39.30 & 0.16 & <0.001\\
            IM10 & 80.49 & 42.41 & 0.07 & <0.001\\
            IM20 & 71.99 & 38.56 & 0.09 & <0.001\\
            IM30 & 64.49 & 35.01 & 0.08 & <0.001\\
            IM40 & 35.96 & 21.90 & 0.10 & <0.001\\
            IM50 & 21.66 & 15.19 & 0.14 & <0.001\\
            IM60 & 9.89 & 9.63 & 0.04 & 0.021\\
            IM70 & 10.74 & 10.02 & 0.05 & 0.007\\
            IM80 & 11.49 & 10.37 & 0.05 & 0.004\\
            IM90 & 21.11 & 14.86 & 0.05 & 0.008\\
            IM100 & 39.95 & 23.59 & 0.10 & <0.001\\
        \end{tabular}
    }
    \caption{\label{tab:corCoef} Comparing cooperation rates with scores earned. Shaded rows indicate that the correlation between cooperation rate and score is negative and statistically significant ($p<0.05$). This table presents results for network ER5/Reward M3.}
\end{table}

\subsection{Moody Conditional Cooperation}

Moody Conditional Cooperation has two requirements: hysteresis (i.e. an agent must be more likely to cooperate if it cooperated on the last turn) and conditionality (i.e. an agent must be more likely to cooperate if its neighbours were predominantly cooperators on the last turn)~\cite{grujic2010social}. Example settings for each of these are given by Table \ref{tab:hyst} and Table \ref{tab:cond} respectively. In the former, we expect to see higher cooperation rates after a previous cooperation than after a defection (with $p<0.05$), which is the case for all BACTD and IM agents, but not for our BACTS agents. In the latter, we expect higher cooperation rates when surrounded by a majority of cooperators rather than defectors (with $p<0.05$), which is the case for some BACTD, some BACTS, and all IM agents. 

\begin{table}[tb]
    \centering{
        \begin{tabular}{c|N|N|N|N}
            Agent &  \mcb{C After}  &  \mcb{C After} & \mcb{G-test} & \mcnb{G-test}  \\
            &  \mcb{C \%}     &  \mcb{D \%}    & \mcb{stat.}     & \mcnb{p-val.} \\
            \hline
            \rowcolor{LightCyan}
            BACTD0 & 68.51 & 31.26 & \mcb{\num{2.69E+4}} & <0.001 \\
            \rowcolor{LightCyan}
            BACTD1 & 64.96 & 29.68 & \mcb{\num{2.53E+4}} & <0.001 \\
            \rowcolor{LightCyan}
            BACTD10 & 60.37 & 24.74 & \mcb{\num{2.54E+4}} & <0.001 \\
            BACTS0 & 69.44 & 69.56 & 0.31 & 0.58 \\
            BACTS1 & 62.89 & 63.42 & 5.67 & 0.017 \\
            BACTS10 & 48.07 & 49.61 & 47.28 & <0.001 \\
            \hline
            \rowcolor{LightCyan}
            IM0 & 95.93 & 13.48 & \mcb{\num{1.35E+5}} & <0.001 \\
            \rowcolor{LightCyan}
            IM10 & 95.33 & 22.77 & \mcb{\num{9.12E+4}} & <0.001 \\
            \rowcolor{LightCyan}
            IM20 & 92.79 & 20.62 & \mcb{\num{1.04E+5}} & <0.001 \\
            \rowcolor{LightCyan}
            IM30 & 86.23 & 26.15 & \mcb{\num{7.42E+4}} & <0.001 \\
            \rowcolor{LightCyan}
            IM40 & 73.77 & 14.30 & \mcb{\num{7.26E+4}} & <0.001 \\
            \rowcolor{LightCyan}
            IM50 & 56.78 & 11.19 & \mcb{\num{3.70E+4}} & <0.001 \\
            \rowcolor{LightCyan}
            IM60 & 43.26 & 5.41 & \mcb{\num{1.97E+4}} & <0.001 \\
            \rowcolor{LightCyan}
            IM70 & 40.84 & 6.33 & \mcb{\num{1.72E+4}} & <0.001 \\
            \rowcolor{LightCyan}
            IM80 & 40.85 & 6.88 & \mcb{\num{1.72E+4}} & <0.001 \\
            \rowcolor{LightCyan}
            IM90 & 44.94 & 14.02 & \mcb{\num{1.72E+4}} & <0.001 \\
            \rowcolor{LightCyan}
            IM100 & 48.76 & 33.81 & \mcb{\num{4.45E+3}} & <0.001 \\
        \end{tabular}
    }
    \caption{\label{tab:hyst} A comparison of cooperation rates after either cooperating or defecting on the last turn. Shaded rows indicate that cooperation is higher after previously cooperating than it is after previously defecting (i.e. hysteresis is observed) and that the difference is statistically significant ($p<0.05$). This table presents results for network ER5 and Reward M3.}
\end{table}

\begin{table}[tb]
    \centering{
        \begin{tabular}{c|N|N|N|N}
            Agent &  \mcb{C Near}  &  \mcb{C Near} & \mcb{G-test}    & \mcnb{G-test}  \\
            &  \mcb{C \%}    &  \mcb{D \%}   & \mcb{stat.}     & \mcnb{p-val.} \\
            \hline
            \rowcolor{LightCyan}
            BACTD0 & 50.49 & 49.12 & 28.77 & <0.001 \\
            \rowcolor{LightCyan}
            BACTD1 & 46.40 & 45.90 & 3.88 & 0.049 \\
            BACTD10 & 37.36 & 38.96 & 35.79 & <0.001 \\
            BACTS0 & 69.49 & 69.33 & 0.25 & 0.62 \\
            \rowcolor{LightCyan}
            BACTS1 & 63.35 & 62.29 & 14.71 & <0.001 \\
            \rowcolor{LightCyan}
            BACTS10 & 49.55 & 48.22 & 28.85 & <0.001\\
            \hline
            \rowcolor{LightCyan}
            IM0 & 91.01 & 16.70 & \mcb{\num{8.15E+4}} & <0.001 \\
            \rowcolor{LightCyan}
            IM10 & 94.57 & 18.88 & \mcb{\num{8.08E+4}} & <0.001 \\
            \rowcolor{LightCyan}
            IM20 & 93.07 & 15.73 & \mcb{\num{1.06E+5}} & <0.001 \\
            \rowcolor{LightCyan}
            IM30 & 88.39 & 17.87 & \mcb{\num{9.14E+4}} & <0.001 \\
            \rowcolor{LightCyan}
            IM40 & 81.65 & 9.99 & \mcb{\num{9.83E+4}} & <0.001 \\
            \rowcolor{LightCyan}
            IM50 & 72.69 & 8.15 & \mcb{\num{5.76E+4}} & <0.001 \\
            \rowcolor{LightCyan}
            IM60 & 67.73 & 4.21 & \mcb{\num{2.88E+4}} & <0.001 \\
            \rowcolor{LightCyan}
            IM70 & 68.14 & 4.44 & \mcb{\num{3.11E+4}} & <0.001 \\
            \rowcolor{LightCyan}
            IM80 & 69.59 & 4.47 & \mcb{\num{3.71E+4}} & <0.001 \\
            \rowcolor{LightCyan}
            IM90 & 75.84 & 8.38 & \mcb{\num{5.79E+4}} & <0.001 \\
            \rowcolor{LightCyan}
            IM100 & 77.40 & 17.68 & \mcb{\num{6.22E+4}} & <0.001 \\
        \end{tabular}
    }
    \caption{\label{tab:cond} A comparison of cooperation rates after being surrounded by a majority of either cooperators or defectors on the last turn. Shaded rows indicate that cooperation is higher near other cooperators than near defectors (i.e. conditionality is observed) and that the difference is statistically significant ($p<0.05$). This table presents results for network ER5 and Reward M3.} 
\end{table}

Though the various BayesACT agents had performed fairly similarly until this point, we see here that those using the default set of initial EPA values seem to display a strong hysteresis (in 100\% of test settings), while those using the study set do not (only 4\% of all test settings). Additionally, even the best of the BACTS agents, BACTS0, showed a hysteresis in only 11\% of network/matrix combinations. We believe that this is most likely a result of the larger difference between the EPAs of the cooperate and defect actions in the default set [\epa{2.1}{1.5}{0.8} and \epa{-2.3}{-0.5}{-0.8} vs \epa{1.4}{0.1}{0.2} and \epa{-0.7}{0.9}{0.70}] resulting in higher deflections and hence more severe reactions.

Looking at the best settings reveals that BACTD0 displays conditionality in 44\% of network/matrix combinations, while BACTS1 does so in 67\% of them. Viewed in aggregate across all reward matrices and all network structures, 22\% of BACTD and 44\% of BACTS agents exhibit statistically significant conditionality. We therefore make only a modest claim of conditional cooperation that we return to in the discussion.

All imitator agents, on the other hand, have both strong hysteresis and conditionality (i.e. 100\% of settings). Given the tendency of these systems to develop towards either high cooperation or total defection, this is not surprising.

\subsection{Player Type Stratification}

According to Gruji\'c et al.~\cite{grujic2010social}, human players can be broadly classified as belonging to one of 5 groups: those who only cooperate, those who mostly cooperate (at least two times in three), moody conditional cooperators, those who mostly defect (at least two times in three), and those who only defect. These groups tend to be unevenly populated, with the most in the middle group, fewer in either of the ``mostly'' groups, and very few in either of the ``pure'' groups.
Thus, we can define a stratification condition as shown in Equation~\ref{eq:strat}.
\begin{equation} \label{eq:strat}
\begin{aligned}
& (Mixed \% & > & Mostly D \% & > & Pure D \% & > & 0) \\
\wedge & (Mixed \% & > & Mostly C \% & > & Pure C \% & > & 0)
\end{aligned}
\end{equation}

For human systems that settle on a comparatively high level of cooperation, it makes sense to consider MCC as the middle group. However, given the imitator agents tendency towards full defection, MCC behaviour is not limited to the middle third of cooperation rates. Hence, we have simply called the middle group ``Mixed'', and dealt with MCC on its own. For one experimental setting, this data is given by Table \ref{tab:strat}, where we have shaded rows that meet the stratification condition.

\begin{table}[tb]
    \small
    \centering{
        \begin{tabular}{c|N|N|N|N|N}
            Agent &  \mcb{Pure}    &  \mcb{Mostly} & \mcb{Mixed}   & \mcb{Mostly}  & \mcnb{Pure} \\
            &  \mcb{D \%}    &  \mcb{D \%}   & \mcb{\%}      & \mcb{C \%}    & \mcnb{C \%} \\
            \hline
            \rowcolor{LightCyan}
            BACTD0 & 5.77 & 28.55 & 31.57 & 28.88 & 5.24\\
            BACTD1 & 5.92 & 33.08 & 32.34 & 25.56 & 3.11\\
            BACTD10 & 6.27 & 43.99 & 28.43 & 20.06 & 1.24\\
            BACTS0 & 0.00 & 0.00 & 33.25 & 66.75 & 0.00\\
            BACTS1 & 0.00 & 0.00 & 72.46 & 27.54 & 0.00\\
            BACTS10 & 0.00 & 0.47 & 99.32 & 0.21 & 0.00\\
            \hline
            IM0 & 11.42 & 7.22 & 4.35 & 60.89 & 16.12\\
            IM10 & 3.55 & 2.57 & 5.83 & 80.74 & 7.31\\
            IM20 & 4.76 & 6.48 & 13.73 & 71.21 & 3.82\\
            IM30 & 4.32 & 10.33 & 25.65 & 58.28 & 1.42\\
            IM40 & 9.35 & 39.91 & 33.58 & 16.69 & 0.47\\
            IM50 & 3.91 & 70.00 & 25.53 & 0.38 & 0.18\\
            IM60 & 8.91 & 87.40 & 3.46 & 0.00 & 0.24\\
            IM70 & 3.34 & 95.21 & 1.18 & 0.00 & 0.27\\
            IM80 & 2.69 & 93.91 & 3.20 & 0.00 & 0.21\\
            IM90 & 0.92 & 84.44 & 9.64 & 4.79 & 0.21\\
            \rowcolor{LightCyan}
            IM100 & 0.24 & 38.67 & 52.25 & 8.73 & 0.12\\
        \end{tabular}
    }
    \caption{\label{tab:strat} A comparison of agent stratification according to the 5 groups of Gruji\'c et al.~\protect\cite{grujic2010social}. Shaded rows indicate adherence to Equation \ref{eq:strat}. This table presents results for network ER5 and Reward M3.}
\end{table}

Averaged across all reward matrices and all network types, we see see 100\% satisfaction of Equation \ref{eq:strat} from BACTD0. There is therefore at least one setting of BayesACT that behaves in a human-like fashion according to Requirement 5 of Gruji\'c et al.~\cite{grujic2010social}. BACTS remains unstratified for all parameter settings, while 3\% of IM agents show strong stratification (i.e. adhere to Equation \ref{eq:strat}).

On the other hand, the best IM bot, IM100, only satisfies Equation \ref{eq:strat} in 33\% of network/matrix combinations. If IM100 is discounted (because $p=100$ corresponds to never actually using strategic imitation, but rather always imitating a neighbour at random), then there are no imitator bots that meet the stratification requirement in any test setting.


\section{Discussion and Future Work}
We have shown how agents based on Affect Control Theory (BayesACT agents) clearly replicate four of the five properties of human play in the iterated networked prisoner's dilemma: network invariance, cooperation-score anti-correlation, and stratification.  We have further shown that these BayesACT agents replicate a fourth property (moody conditional cooperation) in $2/3$ of the cases we have considered. Agents based on the more commonly used homophilic principle of imitation only replicate the moody conditional cooperation property, which they are specifically designed to reproduce \cite{vilone2014social}.

The BayesACT agents considered in this paper are the simplest type, and were implemented with only the default base parameter settings.  For example, a BayesACT agent keeps an EPA distribution for the perception of its opponent. As one action must be chosen to be applied to all partners, we modeled the complete set of them as one general ``opponent''. Both this distribution and the one for self are allowed to change as observations are received. For instance, after its partner defects, a BayesACT agent will view that partner through a more negative EPA. However, this process of change is slow by default. By increasing the value of the corresponding tuning parameter (which was held at its default value for these experiments), larger changes in identity may be allowed after every round, allowing for sooner retributive action. Such an outcome could help these agents to better conform to both the conditionality aspect of MCC and the overall drop in cooperation rate over time observed in human players. Further, BayesACT agents can hold a separate model for each opponent, leading to greater differentiation in behaviour between neighbours.

While the imitation-based agents appear to come up short in all requirements but one, it must be noted here that due to the computational cost of BayesACT, we were not able to test the full space of reward as completely as Vilone et al.~\cite{vilone2014social}. It is therefore possible that an ideal setting for the imitator agents was missed amidst that continuous space.

Looking to future work, we are looking to make comparisons with additional agent models, though this is complicated somewhat by the ill-adaptation of many agents to a networked setting. For example, while a network composed of simple strategies that perform well in a 1-on-1 setting like tit-for-tat or win-stay-lose-shift proceed very rapidly towards either full cooperation or full defection, it is possible that incorporating random elements could produce more human-like behaviour. Similarly, a version of Fehr and Schmidt's inequity aversion model~\cite{fehr1999theory}, which creates agents that seek to punish unjust success, modified for use on a network is of interest.

Additionally, we are interested in applying BayesACT to other types of networks. In particular, scale-free networks, which were examined by Vilone et al.~\cite{vilone2014social}, are highly non-uniform, unlike their grid and ER counterparts, and more closely resemble human social networks~\cite{jackson2008social}. However, given that Gruji\'c et al.~\cite{grujic2014comparative} did not have data from human studies on scale-free networks when making their observations, it is not clear that invariance ought to continue in that domain.

It would also be interesting to adjust the clustering of agents within a network. Currently, BayesACT agents are given random initial EPA distributions, creating a continuum of ``good'' to ``bad'' identities. However, these agents are mixed homogeneously throughout the network. If clusters of nodes were assigned similar identities, it could produce persistent microcosms within a wider network, much as envisioned by Hamilton and Axelrod~\cite{hamilton1981evolution}.

Removing the Broadcast restriction opens up further opportunities for future work. If agents do not act simultaneously, then there is freedom in selecting the ordering of these actions. A simple solution would be to rotate the agents such that each gets a turn, but this is not the only option. Recalling Heise's jury model~\cite{heise2013modeling}, we could allow the agent with the highest personal tension (i.e. the squared difference between that agent's fundamental self-sentiment, and the transient self-sentiment resulting from the last interaction in which it participated) to act next.  This could lead to further epxerimentation on the fragmentation and segregation of agents in a networked environment. 

Finally, we introduce the possibility of applying BayesACT to networks in the wild. Though such networks do not generally explicitly involve the Prisoner's Dilemma, they may contain features that can be cast as normal-form games. When a FaceBook user ignores a friend request, or a GitHub user denies a pull request, there is an implicit defection-like interaction with another agent somewhere in the world. If one supposes that such scenarios come with even a portion of the emotional baggage of a Prisoner's Dilemma, then BayesACT may be able to help explain them.

Understanding the social and emotional forces behind interactions between biological and technological agents is increasingly important in the emerging socio-technical society. Prior research suggests that people care at least as much about maintaining social relationships as they do about striving to maximize personal gains in their transactions with others~\cite{andreoni1993rational, deutsch1958trust, flood1958some}. This makes intuitive sense, since maximizing one's gains depends on sustaining valuable relationships over time. This paper has investigated a promising social-psychological model of affective relationships in a constrained and artificial simulation of a social dilemma in an attempt to move towards more affectively and socially aligned technological agents that will facilitate these relationships between biological and technological agents. 

\section{Concluding Remarks}
We have shown that, compared to imitation-based agents, agents based on the social-psychological affect control theory (BayesACT agents) display as emergent properties more of the human qualities identified by Gruji\'c et al.~\cite{grujic2014comparative} in the Iterated Networked Prisoner's Dilemma (INPD). In particular, we show how the observed human behaviours of network structure invariance, anti-correlation of cooperation and reward, player type stratification, and (to a certain extent) moody conditional cooperation, are all emergent properties of these agents. We therefore believe that this work brings us a step closer to reproducing human behaviour in the INPD, and may find application both in domains that require human-like behaviour, and those that wish to probe human reasoning.

\bibliographystyle{abbrv}

\bibliography{cogSci2017References}

\begin{thebibliography}{10}

\bibitem{andreoni1993rational}
J.~Andreoni and J.~H. Miller.
\newblock Rational cooperation in the finitely repeated prisoner's dilemma:
  Experimental evidence.
\newblock {\em The economic journal}, 103(418):570--585, 1993.

\bibitem{asghar2015intelligent}
N.~Asghar and J.~Hoey.
\newblock Intelligent affect: Rational decision making for socially aligned
  agents.
\newblock In {\em Proceedings of the 31st Conference on Uncertainty in
  Artificial Intelligence}, pages 12--16, 2015.

\bibitem{Damasio1994}
A.~R. Damasio.
\newblock {\em Descartes' error: Emotion, reason, and the human brain}.
\newblock Putnam's sons, 1994.

\bibitem{deutsch1958trust}
M.~Deutsch.
\newblock Trust and suspicion.
\newblock {\em Journal of conflict resolution}, pages 265--279, 1958.

\bibitem{fehr1999theory}
E.~Fehr and K.~M. Schmidt.
\newblock A theory of fairness, competition, and cooperation.
\newblock {\em The quarterly journal of economics}, 114(3):817--868, 1999.

\bibitem{flood1958some}
M.~M. Flood.
\newblock Some experimental games.
\newblock {\em Management Science}, 5(1):5--26, 1958.

\bibitem{Fontaine2007}
J.~R.~J. Fontaine, K.~R. Scherer, E.~B. Roesch, and P.~C. Ellsworth.
\newblock The world of emotions is not two-dimensional.
\newblock {\em Psychological Science}, 18:1050 -- 1057, 2007.

\bibitem{francis2006mean}
C.~Francis and D.~R. Heise.
\newblock Mean affective ratings of 1,500 concepts by {I}ndiana {U}niversity
  undergraduates in 2002-3.
\newblock {\em Data in Computer Program Interact}, 2006.

\bibitem{gracia2012heterogeneous}
C.~Gracia-L{\'a}zaro, A.~Ferrer, G.~Ruiz, A.~Taranc{\'o}n, J.~A. Cuesta,
  A.~S{\'a}nchez, and Y.~Moreno.
\newblock Heterogeneous networks do not promote cooperation when humans play a
  prisoner’s dilemma.
\newblock {\em Proceedings of the National Academy of Sciences},
  109(32):12922--12926, 2012.

\bibitem{grujic2010social}
J.~Gruji{\'c}, C.~Fosco, L.~Araujo, J.~A. Cuesta, and A.~S{\'a}nchez.
\newblock Social experiments in the mesoscale: Humans playing a spatial
  prisoner's dilemma.
\newblock {\em PloS one}, 5(11):e13749, 2010.

\bibitem{grujic2014comparative}
J.~Gruji{\'c}, C.~Gracia-L{\'a}zaro, M.~Milinski, D.~Semmann, A.~Traulsen,
  J.~A. Cuesta, Y.~Moreno, and A.~S{\'a}nchez.
\newblock A comparative analysis of spatial prisoner's dilemma experiments:
  Conditional cooperation and payoff irrelevance.
\newblock {\em Scientific reports}, 4:4615, 2014.

\bibitem{grujic2012consistent}
J.~Gruji{\'c}, T.~R{\"o}hl, D.~Semmann, M.~Milinski, and A.~Traulsen.
\newblock Consistent strategy updating in spatial and non-spatial behavioral
  experiments does not promote cooperation in social networks.
\newblock {\em PLoS One}, 7(11):e47718, 2012.

\bibitem{hamilton1981evolution}
W.~D. Hamilton and R.~Axelrod.
\newblock The evolution of cooperation.
\newblock {\em Science}, 211(27):1390--1396, 1981.

\bibitem{heise1977social}
D.~R. Heise.
\newblock Social action as the control of affect.
\newblock {\em Behavioral Science}, 22(3):163--177, 1977.

\bibitem{Heise2007}
D.~R. Heise.
\newblock {\em Expressive Order: Confirming Sentiments in Social Actions}.
\newblock Springer, 2007.

\bibitem{heise2010surveying}
D.~R. Heise.
\newblock {\em Surveying cultures: Discovering shared conceptions and
  sentiments}.
\newblock John Wiley \& Sons, 2010.

\bibitem{Heise2013}
D.~R. Heise.
\newblock Modeling interactions in small groups.
\newblock {\em Social Psychology Quarterly}, 76:52--72, 2013.

\bibitem{heise2013modeling}
D.~R. Heise.
\newblock Modeling interactions in small groups.
\newblock {\em Social Psychology Quarterly}, page 0190272512467654, 2013.

\bibitem{hoey2013bayesian}
J.~Hoey, T.~Schroder, and A.~Alhothali.
\newblock Bayesian affect control theory.
\newblock In {\em Affective Computing and Intelligent Interaction (ACII), 2013
  Humaine Association Conference on}, pages 166--172. IEEE, 2013.

\bibitem{HoeyBACT15}
J.~Hoey, T.~Schr\"{o}der, and A.~Alhothali.
\newblock Affect control processes: Intelligent affective interaction using a
  partially observable {M}arkov decision process.
\newblock {\em Artificial Intelligence}, 230:134--172, January 2016.

\bibitem{jackson2008social}
M.~O. Jackson.
\newblock {\em Social and Economic Networks}, volume~3.
\newblock Princeton University Press, 2008.

\bibitem{jung2016grounding}
J.~D. Jung, J.~Hoey, J.~H. Morgan, T.~Schr{\"o}der, and I.~Wolf.
\newblock Grounding social interaction with affective intelligence.
\newblock In {\em Canadian Conference on Artificial Intelligence}, pages
  52--57. Springer, 2016.

\bibitem{Kahneman11}
D.~Kahneman.
\newblock {\em Thinking, Fast and Slow}.
\newblock Doubleday, 2011.

\bibitem{kiesler1996prisoner}
S.~Kiesler, L.~Sproull, and K.~Waters.
\newblock A prisoner's dilemma experiment on cooperation with people and
  human-like computers.
\newblock {\em Journal of personality and social psychology}, 70(1):47, 1996.

\bibitem{kollock1998social}
P.~Kollock.
\newblock Social dilemmas: The anatomy of cooperation.
\newblock {\em Annual review of sociology}, pages 183--214, 1998.

\bibitem{kreps1982rational}
D.~M. Kreps, P.~Milgrom, J.~Roberts, and R.~Wilson.
\newblock Rational cooperation in the finitely-repeated prisoners' dilemma.
\newblock Technical report, DTIC Document, 1982.

\bibitem{MacKinnon2014}
N.~J. MacKinnnon and D.~T. Robinson.
\newblock 25 years of research in affect control theory.
\newblock {\em Advances in Group Processing}, 31, 2014.

\bibitem{mcdonald2014multiple}
J.~McDonald.
\newblock Multiple comparisons: Controlling the false discovery rate:
  Benjamini--hochberg procedure.
\newblock {\em Handbook of Biological Statistics. Baltimore, Maryland: Sparkly
  House Publishing}, pages 254--60, 2014.

\bibitem{nowak2006five}
M.~A. Nowak.
\newblock Five rules for the evolution of cooperation.
\newblock {\em science}, 314(5805):1560--1563, 2006.

\bibitem{osgood1952nature}
C.~E. Osgood.
\newblock The nature and measurement of meaning.
\newblock {\em Psychological bulletin}, 49(3):197, 1952.

\bibitem{Osgood1975}
C.~E. Osgood, W.~H. May, and M.~S. Miron.
\newblock {\em Cross-Cultural Universals of Affective Meaning}.
\newblock University of Illinois Press, 1975.

\bibitem{osgood1975cross}
C.~E. Osgood, W.~H. May, and M.~S. Miron.
\newblock {\em Cross-cultural universals of affective meaning}.
\newblock University of Illinois Press, 1975.

\bibitem{rand2011dynamic}
D.~G. Rand, S.~Arbesman, and N.~A. Christakis.
\newblock Dynamic social networks promote cooperation in experiments with
  humans.
\newblock {\em Proceedings of the National Academy of Sciences},
  108(48):19193--19198, 2011.

\bibitem{Scholl2013}
W.~Scholl.
\newblock The socio-emotional basis of human interaction and communication: How
  we construct our social world.
\newblock {\em Social Science Information}, 52:3 -- 33, 2013.

\bibitem{SchroederHoeyRogers2016}
T.~Schr\"{o}der, J.~Hoey, and K.~B. Rogers.
\newblock Modeling dynamic identities and uncertainty in social interactions:
  Bayesian affect control theory.
\newblock {\em American Sociological Review}, 81(4), 2016.

\bibitem{Schroeder2013}
T.~Schr\"{o}der, J.~Netzel, C.~Schermuly, and W.~Scholl.
\newblock Culture-constrained affective consistency of interpersonal behavior:
  A test of affect control theory with nonverbal expressions.
\newblock {\em Social Psychology}, 44:47--58, 2013.

\bibitem{Schroeder2009}
T.~Schr\"{o}der and W.~Scholl.
\newblock Affective dynamics of leadership: An experimental test of affect
  control theory.
\newblock {\em Social Psychology Quarterly}, 72:180--197, 2009.

\bibitem{silver2010monte}
D.~Silver and J.~Veness.
\newblock {M}onte-{C}arlo planning in large {POMDP}s.
\newblock In {\em Advances in neural information processing systems}, pages
  2164--2172, 2010.

\bibitem{SmithLovin2007}
L.~Smith-Lovin.
\newblock The strength of weak identities: Social structural sources of self,
  situation and emotional experience.
\newblock {\em Social Psychology Quarterly}, 70(2):106--124, 2007.

\bibitem{trivers1971evolution}
R.~L. Trivers.
\newblock The evolution of reciprocal altruism.
\newblock {\em Quarterly review of biology}, pages 35--57, 1971.

\bibitem{vilone2014social}
D.~Vilone, J.~J. Ramasco, A.~S{\'a}nchez, and M.~San~Miguel.
\newblock Social imitation versus strategic choice, or consensus versus
  cooperation, in the networked prisoner's dilemma.
\newblock {\em Physical Review E}, 90(2):022810, 2014.

\end{thebibliography}

\end{document}